\documentclass[pra,twocolumn
,showpacs,preprintnumbers,amsmath,amssymb]{revtex4}
\usepackage{graphicx}
\usepackage{dcolumn}
\usepackage{bm}
\newcommand{\p}{\partial}
\newcommand{\vek}[1]{\bm{\mathrm{#1}}}
\newcommand{\sst}[1]{\scriptstyle{#1}}

\newcommand{\Eq}[1]{Eq.\@ (\ref{#1})}
\newcommand{\Eqs}[1]{Eqs.\@ (\ref{#1})}

\begin{document}

\title{Dipole Oscillations in Fermionic Mixtures}

\author{S. Chiacchiera}
\affiliation{Centro de F\'{i}sica Computacional, Department of 
Physics, University of Coimbra, P-3004-516, Coimbra, Portugal}

\author{T. Macr\`{i}}
\affiliation{SISSA and INFN, Sezione di Trieste, via Beirut 2/4, I-34151, 
Trieste, Italy}

\author{A. Trombettoni}
\affiliation{SISSA and INFN, Sezione di Trieste, via Beirut 2/4, I-34151, 
Trieste, Italy}


\begin{abstract}
We study dipole oscillations in a general fermionic mixture: 
starting from the Boltzmann equation, we classify the different solutions 
in the parameter space through the number of real eigenvalues of the small 
oscillations matrix. We discuss how this number can be computed using the Sturm
algorithm and its relation with the properties
of the Laplace transform of the experimental quantities. 
After considering two components in harmonic potentials 
having different trapping frequencies, we study dipole oscillations in 
three-component mixtures. 
Explicit computations are done for realistic experimental setups using
the classical Boltzmann equation without intra-species interactions.
A brief discussion of the application of this
classification to general collective oscillations is also presented.  
\end{abstract}
\maketitle

\section{Introduction}

The study of collective modes is a major tool to characterize and unravel 
the effects of interparticle interactions in a broad range of physical systems 
and in particular in experiments of ultracold atoms, where the scattering 
length can be tuned through Feshbach resonances \cite{PITAEVSKII03,PETHICK08}. 
For ultracold bosons the collective oscillation
frequencies in single-component gases depend 
on the interaction energy and in multi-component gases 
they depend on both intra- and
inter-species interactions; 
for ultracold fermions the $s$-wave intra-species 
scattering length vanishes due to Pauli 
principle and the collective modes depend on the interactions between the 
different components, so that their study is crucial in 
multi-component fermionic mixtures. 

In the last decade, the impressive experimental progresses in trapping and 
controlling Fermi gases
\cite{KETTERLE08,GIORGINI08} allowed their study in highly controllable setups: 
it became possible to lower
the temperature to reveal degeneracy effects \cite{DEMARCO99}, 
tune the interactions to explore the BEC-BCS crossover 
\cite{REGAL03,JOCHIM03,ZWIERLEIN03,BOURDEL04,PARTRIDGE05,KINAST05,VEERAVALLI08}, 
superimpose optical lattices \cite{MODUGNO03,CHIN06,STOFERLE06,SCHNEIDER08} 
and polarize two-component Fermi mixtures 
\cite{ZWIERLEIN06,PARTRIDGE06}. Collective modes 
in two-component mixtures have also been experimentally studied 
\cite{GENSEMER01,KINAST04,BARTENSTEIN04,WRIGHT07,RIEDL09,NASCIMBENE09}.

Among these progresses, 
it recently became experimentally manageable to study three-component 
fermionic mixtures of $^{6}Li$ \cite{OTTENSTEIN08} and 
of $^{6}Li-^{40}K$ \cite{WILLE08,SPIEGELHADER09}. 
In \cite{OTTENSTEIN08} a degenerate Fermi gas consisting of 
three different hyperfine states of $^{6}Li$ was created: the three 
scattering lengths $a_{12}$, $a_{13}$ and $a_{23}$ are different and tunable; 
the collisional stability of the gas has been studied, showing 
rather long lifetimes when one scattering length is close 
to the unitary limit and the other two are small \cite{OTTENSTEIN08}. 
In \cite{WILLE08} the interspecies Feshbach resonances of a mixture 
of $^{6}Li$ and $^{40}K$ were studied: the collisional stability 
has been recently determined 
\cite{SPIEGELHADER09}, showing that a small sample 
of $^{40}K$ immersed in a two-component $^6Li$ mixture is stable with 
very low loss rates
for large negative scattering lengths between the two $^6Li$. 

These systems are particularly promising for several reasons: 
they offer a way to investigate new strongly 
interacting Fermi systems; thanks to their enlarged parameter space, 
they could present novel phases 
related to the formation of unconventional pairing and stability of trions  
\cite{RAPP07,CAPPONI08,GUAN08,SILVA09,BEDAQUEA09}; 
moreover, the possibility of having many components with tunable interactions 
can be used to study mechanisms of color superconductivity \cite{ALFORD08}  
and, in perspective, to simulate parts of the QCD phase diagram 
\cite{RAJAGOPAL00}.

A natural way to characterize multi-component fermionic mixtures is provided 
by the study of their modes. 
For two-component Fermi mixtures, 
collective oscillations have been investigated 
in several situations of experimental relevance 
\cite{VICHI99,VICHI00,BRUUN01,MARUYAMA06,BRUUN07,LAZARIDES08,DAHAL08,RIEDL09,CHIACCHIERA09} 
(see more references in the reviews 
\cite{KETTERLE08,GIORGINI08}) and their 
behaviour studied when different parameters are varied: 
temperature, scattering length, 
polarization (i.e., the relative number of atoms of the components) 
and mass ratio. The variety of possible dynamical regimes grows 
when the number of components increases: 
quantized vortices of three-component Fermi mixtures with attractive 
interactions have been recently studied \cite{CATELANI09}. 

The goal of the present paper is to discuss a simple way 
to characterize 
the properties of the collective oscillations of 
a general multi-component mixture. 
Restricting to temperatures 
larger than the superfluid critical temperature 
(see discussion in Section II),  
we will describe the Fermi gas mixture using the Boltzmann equation, 
with no intra-species interaction. 
Introducing a Gaussian ansatz for the distribution function, 
the equations for the small oscillations are determined: 
a convenient way to classify the different regimes in the parameter space is based on 
the study of the eigenvalues of the small oscillations matrix. Once that the eigenvalues are found, 
the experimental quantities 
(like the center of mass positions for dipole oscillations) 
can be determined. The real 
part of the eigenvalues determines the damping of the quantities, while the imaginary part 
is related to the oscillations: we show that the number of peaks of the 
analytical continuation of the Laplace transform of the physical quantities gives 
the number of real solutions (at variance, the peaks of the 
Fourier transform do not give the number 
of complex solutions). 
One can then classify the solutions of the linearized Boltzmann equation 
through the number of real eigenvalues of the small 
oscillations matrix. 
A real root corresponds to an overdamped eigenmode, whereas complex
ones correspond to oscillatory  modes and purely imaginary to undamped 
oscillatory modes.
The small oscillation matrix is real, therefore complex eigenvalues
appear in pairs of complex conjugates and the
number of oscillation frequencies 
is simply the number of pairs of complex roots.
Counting the overdamped eigenmodes (i.e. the real roots), 
one has a measure of the degree of collectivity
of the system: the larger this number, the stronger the 
effect of interactions and trap confinement in reducing the  
degrees of freedom of the mixture.
We will show that this approach is complementary 
to the study of the oscillatory frequencies, 
and convenient to classify the properties of the small oscillations solutions.
We will focus on the case of dipole oscillations
(i.e, one-dimensional out-of-phase motion of the centers of mass of the various
components), 
but the proposed classification works 
for general collective modes of multi-component fermionic mixtures.

The plan of the paper is the following: 
In the next Section we review the formalism of the Boltzmann equation 
and the method of averages, applying it to a multi-component fermionic mixture 
and deriving the equations of motion for the centers of mass 
of the components. In Section III we study the dipole oscillations of 
a two-component mixture in harmonic potentials 
having different trapping frequencies: 
indeed, when the trapping frequencies are equal, 
two of the eigenvalues of the $4 \times 4$ 
small oscillation matrix are purely imaginary 
due to the Kohn theorem \cite{KOHN61,PITAEVSKII03,PETHICK08}, 
while for different trapping frequencies one can have either
$0$ or $2$ or $4$ real solutions, similarly to what happens 
for a three-component mixture with equal frequencies. 
In Section III we also recall that a convenient way to count the 
number of real eigenvalues is based on the 
Sturm theorem and we discuss how to characterize the different 
solutions starting from the experimental quantities. In Section IV 
we study dipole oscillations in 
three-component mixtures, presenting results for a mixture of 
three hyperfine levels of $^{6}Li$ and for a mixture of 
two hyperfine levels of $^{6}Li$  and 
one of $^{40}K$, like in the recently reported experimental results 
\cite{OTTENSTEIN08,WILLE08,SPIEGELHADER09}. 
In Sec. V we draw our conclusions and discuss possible further developments, 
whereas in the Appendices additional material is presented.

\section{Boltzmann equation for a multi-component mixture}

In this Section we discuss the main properties of the Boltzmann equation for a 
trapped multi-component fermionic mixture.
Consider a trapped balanced two-component Fermi gas; 
when the temperature $T$ is significantly 
larger than the Fermi temperature $T_F$, 
the gas behaves classically 
and is in the collisionless regime.
Decreasing $T$, 
the frequency of collisions increases leading to
a hydrodynamical behaviour.
Further decreasing the temperature $T$
(but with $T$ larger than the superfluid critical temperature), 
degeneracy effects become strong and one would expect the
mixture to be collisionless again. 
For weak (inter-species) interactions, this sequence 
collisionless-hydrodynamical-collisionless 
actually occurs as $T/T_F$ decreases from values 
larger than one towards smaller ones.
However, if the interactions are strong (i.e., close to the unitary limit) 
the sequence found in the experiments \cite{WRIGHT07,RIEDL09} is 
collisionless-hydrodynamical as it would be in a classical gas 
(i.e, without degeneracy effects): the gas remains hydrodynamical. 
This finding can be explained by taking into account in-medium effects: 
the in-medium enhancement of the cross-section 
compensates Pauli blocking at low temperature 
leading to a hydrodynamical behaviour \cite{RIEDL09,CHIACCHIERA09}. 
In other words, 
classical statistics seems to work well when interactions are strong 
because Pauli blocking and in-medium 
effects almost cancel each other.
In summary, depending on the interaction strength,
either degeneracy dominates or is compensated 
by medium effects and the low temperature behaviour is different in the two cases;
however, whatever the interaction strength is, for temperatures above $\sim 0.5T_F$  
the Boltzmann equation framework 
with no intra-species interactions and classical collision term 
with no medium-effects for the inter-species interactions is found to work quite well, 
not only qualitatively, 
in reproducing the collective modes
\cite{RIEDL09,CHIACCHIERA09}.

We will take advantage of this finding,
and describe the Fermi nature
of the particles simply by forbidding intra-species interactions. 
In multi-component imbalanced mixtures each component has its own Fermi temperature
$T_{F,\alpha}$,
therefore we expect our scheme to work well above $\sim 0.5 \max\{T_{F,\alpha}\}$.

If we restrict ourselves to the normal phase 
(that is, for temperatures above the critical temperature for superfluidity), 
each component of the mixture can be 
described by a semiclassical distribution function 
$f_{\alpha}=f_{\alpha}(\vek{r},\vek{p},t)$ 
obeying the Boltzmann equation \cite{UHLENBECK63}.
Here and in the following 
we denote the $N$ components of the multi-component mixture 
by $\alpha=1,\cdots,N$, the $\alpha$-th 
component having $N_\alpha$ atoms of mass $m_\alpha$. 
The function ${f}_{\alpha}(\vek{r},\vek{p},t)$ 
gives the probability of finding a particle of species $\alpha$
at time $t$ in the $d^3rd^3p$ volume centered in 
($\vek{r},\vek{p}$) in phase space. The normalization condition of $ f_{\alpha}$ is
\begin{equation}\label{eq:normf}
\int d\Gamma f_{\alpha}(\vek{r},\vek{p},t)= N_\alpha~,
\quad \textrm{with}\quad d\Gamma \equiv d^3rd^3p ~,
\end{equation}
and the density is 
\begin{equation}
\rho_{\alpha}(\vek{r},t) =\int d^3 p f_{\alpha}(\vek{r},\vek{p},t)~.
\end{equation}

The Boltzmann equation for our multi-component mixture reads 
\begin{equation}\label{eq:multiBoltzmann}
\frac{\partial f_\alpha}{\partial t}
+\dot{\vek{r}}_{\alpha}\cdot\frac{\p f_{\alpha}}{\p \vek{r}} 
+ \dot{\vek{p}}_{\alpha}\cdot \frac{\p f_{\alpha}}{\p \vek{p}} = 
-{\sum^N_{\gamma=1}}{'} I_{\alpha\gamma}  
\end{equation}
where 
\begin{equation}
\dot{\vek{p}}_{\alpha}=-\frac{\p {\cal V}_{\alpha}}{\p \vek{r}},
\quad \dot{\vek{r}}_{\alpha}=\frac{\vek{p}}{m_\alpha}~.
\end{equation}
The right-hand side of Boltzmann equation 
(\ref{eq:multiBoltzmann}) 
is a sum of collision integrals and the $'$ on the sum means that 
we sum on all $\gamma \neq \alpha$. 
${\cal V}_{\alpha}$ is the single particle potential felt by the $\alpha$ species. In
general, different components feel different anisotropic trapping 
potentials: 
\begin{equation}
{\cal V}_\alpha(\vek{r})\equiv\frac{m_\alpha}{2} \left ( \omega_{\alpha x}^2 x^2 
+\omega_{\alpha y}^2 y^2 +\omega_{\alpha z}^2 z^2 \right)
\label{potentials}
\end{equation}
(we will not consider in the following
the case in which the harmonic potentials ${\cal V}_\alpha$ have different centers 
\cite{WILLIAMS00}). 

The collision integrals for classical statistics read \cite{UHLENBECK63}
\begin{equation}\label{eq:collisionintegral}
I_{\alpha\gamma}  \equiv \int d^3p_1 
d\Omega \frac{d\sigma_{\alpha\gamma}}{d\Omega}
\left|\frac{\vek{p}}{m_\alpha}-\frac{\vek{p}_1}{m_\gamma}\right|
\left(f_{\alpha}f_{\gamma1}
-f_{\alpha}^\prime f_{\gamma1}^\prime\right)
\end{equation}
and represent the variation of $f_{\alpha}$ due to collisions 
with particles of type $\gamma$.
In the latter $\frac{d\sigma_{\alpha\gamma}}{d\Omega}$ is the differential cross 
section of an $\alpha$ atom and 
a $\gamma$ atom in the center of mass of the collision and $\Omega$ is the
angle between the relative outgoing and relative ingoing momenta
 of the two colliding 
particles (in our model there is no intra-species interaction: 
$d\sigma_{\alpha\alpha}=0$).
The atoms $\alpha$ and $\gamma$ have 
respectively momenta $\vek{p}$ and $\vek{p}_1$ 
before the scattering and $\vek{p}^\prime$ and $\vek{p}_1^\prime$ after: 
momentum and kinetic energy are conserved in the collision 
[$\vek{p}+\vek{p_1}=
\vek{p}^\prime+\vek{p}_1^\prime$ and $p^2/2m_\alpha+p_1^2/2m_\gamma=
p^{\prime 2} /2m_\alpha+p_1^{\prime 2}/2m_\gamma$] and all the $f$ are evaluated at the same point 
and time, but different momenta: $f_{\alpha}=f_{\alpha}(\vek{r},\vek{p},t)$, 
$f_{\gamma1}=f_{\gamma}(\vek{r},\vek{p}_1,t)$, 
$f_{\alpha}^\prime=f_{\alpha}(\vek{r},\vek{p}^\prime,t)$ and 
$f_{\gamma1}^\prime=f_{\gamma}(\vek{r},\vek{p}_1^\prime,t)$.

With the classical collision term given in Eq. (\ref{eq:collisionintegral}) 
the equilibrium distribution function is the Maxwell-Boltzmann 
distribution 
\begin{equation}\label{eq:fbar}
\bar{f}_{\alpha}(\vek{r},\vek{p})=\frac{\exp{\{-\beta[p^2/(2m_\alpha)+
{\cal V}_\alpha(\vek{r})-\mu_\alpha]}\}}{(2\pi \hbar)^3}~,
\end{equation}
where $\beta\equiv 1/k_BT$ and the chemical potential is fixed by
the normalization condition \Eq{eq:normf}. 
For the harmonic trapping potential of \Eq{potentials} it is 
\begin{equation}
\mu_\alpha= \frac{1}{\beta}\ln[(\beta \bar{\omega}_\alpha)^3N_\alpha]~,
\end{equation}
where 
\begin{equation}
\bar{\omega}_\alpha\equiv(\omega_{\alpha x} \omega_{\alpha y} \omega_{\alpha z})^{1/3}
\label{omega_bar} 
\end{equation}
is the trap average frequency. 

The effect of Fermi or Bose statistics (blocking or 
anti-blocking of the final state of a collision) 
can be incorporated in the collision integral by adding appropriate factors: 
then, consistently, the equilibrium distribution
function to which the system is driven by collisions is a Fermi or Bose one \cite{UHLENBECK63}.

To study collective oscillations without directly solving the Boltzmann equation, it is useful to derive equations 
for integrated average quantities: defining the average of 
a generic quantity $\chi=\chi(\vek{r},\vek{p})$ in the component $\alpha$ as
\begin{equation}
\left\langle \chi \right\rangle_\alpha \equiv \frac{1}{N_\alpha}\int d\Gamma f_{\alpha}(\vek{r},\vek{p},t) \chi(\vek{r},\vek{p})~,
\end{equation}
from the Boltzmann equation (\ref{eq:multiBoltzmann}) one finds
\begin{equation}\label{eq:evoave}
\frac{d \left\langle \chi \right\rangle_\alpha}{d t}-
\left\langle \frac{\vek p}{m_\alpha}\cdot
\frac{\p \chi}{\p \vek{r}} \right\rangle_\alpha
+\left\langle \frac{\p {\cal V}_\alpha}{\p \vek{r}}\cdot\frac{\p \chi}{\p \vek{p}}\right\rangle_\alpha =
-{\sum_{\gamma=1}^N}{'}\left\langle I_{\alpha\gamma}\chi\right\rangle_{\alpha}~,
\end{equation}
where we have defined the collisional average
\begin{equation}\label{eq:coll_aver}
\left\langle I_{\alpha\gamma}\chi\right\rangle_{\alpha}\equiv \frac{1}{N_\alpha}\int d\Gamma 
I_{\alpha\gamma}\chi(\vek{r},\vek{p})~.
\end{equation} 

The collisional averages of a quantity $\chi$ 
satisfy the following constraint in each pair of components
\begin{eqnarray*}
&&N_\alpha \left\langle I_{\alpha\gamma}\chi\right\rangle_{\alpha}
+N_\gamma \left\langle I_{\gamma\alpha}\chi\right\rangle_{\gamma}=
\frac{1}{2}
\int d\Gamma d^3p_1d\Omega
\frac{d\sigma_{\alpha\gamma}}{d\Omega}\\
&&
\times\left|\frac{\vek{p}}{m_\alpha}-\frac{\vek{p}_1}{m_\gamma}\right|\left(f_{\alpha}
f_{\gamma1}
-f_{\alpha}^\prime f_{\gamma1}^\prime\right)\Delta\chi~,
\end{eqnarray*}
where $\Delta\chi\equiv \chi_{\alpha}+\chi_{\gamma1}-\chi^\prime_{\alpha}-
\chi^\prime_{\gamma1}$: summing the latter over 
$\alpha,\gamma$ one finds the following global constraint
\begin{eqnarray}\label{eq:sumIchi}
&&{\sum_{\alpha,\gamma}}{'} N_\alpha \left\langle I_{\alpha\gamma}\chi\right\rangle_{\alpha}=
\frac{1}{4}
{\sum_{\alpha,\gamma}}{'} \int d\Gamma d^3p_1d\Omega
\frac{d\sigma_{\alpha\gamma}}{d\Omega}
\nonumber\\
&&
\times\left|\frac{\vek{p}}{m_\alpha}-\frac{\vek{p}_1}{m_\gamma}\right|\left(f_{\alpha}
f_{\gamma1}
-f_{\alpha}^\prime f_{\gamma1}^\prime\right)\Delta\chi~.
\end{eqnarray}
If $\chi$ is a collisional invariant, i.e., 
if $\Delta\chi=0$ \cite{UHLENBECK63}, then 
\begin{equation}
{\sum_{\alpha,\gamma}}{'} N_\alpha \left\langle I_{\alpha\gamma}\chi\right\rangle_{\alpha}=0~.
\end{equation}

Eq. (\ref{eq:evoave}) gives 
rise to a set of equations involving different observables: 
the goal is to have a closed set of equations to solve. 
This approach has been used to determine in a classical gas  
the shift of the collective frequencies due to the interparticle collisions in a Bose-Einstein condensate above the 
condensation temperature \cite{GUERY99,ALKHAWAJA00,PEDRI03,MULLIN06} and it has been applied to the study 
of the scissor mode in 
two-component Fermi mixtures \cite{BRUUN07} as well as to Fermi-Bose mixtures 
\cite{FERRARI02,FERLAINO03}. 
A convenient way to evaluate the 
collisional averages to close the set of equations 
and study the small oscillations for the linearized Boltzmann equation 
is to make a Gaussian ansatz for the time-dependent distribution function $f$: 
in \cite{GUERY99} the frequency and the damping of quadrupole oscillation in a classical gas 
obtained using the Gaussian ansatz was compared with the numerical solution of the Boltzmann equation showing a 
good agreement.
 
For dipole oscillations 
along the direction $x$, choosing $\chi=x, p_x$ from Eq. (\ref{eq:evoave}) one gets the following $2N$ equations
\begin{equation}
\left\{
\begin{array}{l} \label{eq:twocomp}
\displaystyle{\frac{d\left\langle x \right\rangle_\alpha}{d t}-\frac{\left\langle p_x\right\rangle_\alpha}{m_\alpha} =
- {\sum^N_{\gamma=1}}{'} \left\langle I_{\alpha\gamma} x \right\rangle_\alpha}
\\
\displaystyle{\frac{d\left\langle p_x \right\rangle_\alpha}{d t}+m_\alpha \omega_{\alpha x}^2 \left\langle x\right\rangle_\alpha =
- {\sum^N_{\gamma=1}}{'} \left\langle I_{\alpha\gamma} p_x \right\rangle_\alpha~, }
\end{array}
\right.
\end{equation}
with $\alpha=1,\cdots, N$. 
To close the set of \Eqs{eq:twocomp}, 
adapting the method of \cite{GUERY99} we make the following Gaussian ansatz for the distribution function:
\begin{equation}\label{eq:gaussianansatz}
{f}_{\alpha}(\vek{r},\vek{p},t)=\frac{\exp{
\left\{-\beta\left[\frac{(\vek{p}-\vek{p}_{\alpha})^2}
{2m_\alpha}+{\cal V}_\alpha \left(\vek{r}-\vek{r}_{\alpha} \right)-\mu_\alpha\right]\right\}}}{(2\pi \hbar)^3}~,
\end{equation}
where $\vek{p}_{\alpha}=\vek{p}_{\alpha}(t)$ and 
$\vek{r}_{\alpha}=\vek{r}_{\alpha}(t)$ depend on time. 
Eq. (\ref{eq:gaussianansatz}) is a local equilibrium ansatz: 
it generalizes (\ref{eq:fbar}) by giving it a 
time-dependent average coordinate [$\left\langle\vek{r}\right\rangle_\alpha=\vek{r}_{\alpha}(t)$] 
and average momentum  
[$\left\langle\vek{p}\right\rangle_\alpha=\vek{p}_{\alpha}(t)$];
moreover, it preserves the normalization condition 
($\int d\Gamma f_{\alpha}(\vek{r},\vek{p},t)=N_\alpha$). 

Using the Gaussian ansatz (\ref{eq:gaussianansatz}) 
and linearizing the \Eqs{eq:twocomp} 
(i.e., retaining only terms that are at most linear in 
$\left\langle x \right\rangle_{\alpha}$ and 
$\left\langle p_x \right\rangle_{\alpha}$), one finds a closed set of equations. 
Indeed the collisional averages in the right-hand side of 
\Eqs{eq:twocomp} can be evaluated giving
\begin{subequations}\label{eq:xpxcollave}
\begin{eqnarray}
&&\left\langle I_{\alpha\gamma}x \right\rangle_{\alpha}=0\label{eq:xcollave}\\
&&\left\langle I_{\alpha\gamma}p_x \right\rangle_{\alpha}=
\frac{\mu_{\alpha\gamma}}{\tau_{\alpha\gamma}} N_\gamma 
\left( \frac{\left\langle p_x \right\rangle_\alpha}{m_\alpha} -
\frac{\left\langle p_x \right\rangle_\gamma}{m_\gamma}\right)\label{eq:pxcollave}
\end{eqnarray}
\end{subequations}
where $\mu_{\alpha\gamma}$ is the two body reduced mass 
($\mu_{\alpha\gamma}\equiv m_\alpha m_\gamma / m_{\alpha\gamma}$, 
with $m_{\alpha\gamma}\equiv m_\alpha+m_\gamma$): 
the explicit computation is reviewed in Appendix A. 
In Eq. (\ref{eq:pxcollave}) $\tau_{\alpha\gamma}$ is 
a parameter 
related to the collisions between atoms $\alpha$ and 
$\gamma$ and will turn out to be proportional
to the relaxation time of the dipole mode:  $\tau_{\alpha\gamma}$ is defined as
\begin{equation} \label{eq:tau}
\frac{1}{\tau_{\alpha\gamma}}\equiv\frac{4\beta^2 \hbar^2}{3\pi}
\left(\frac{\bar{\omega}_\alpha\bar{\omega}_\gamma}
{\bar{\omega}_{\alpha\gamma}}
\right)^3 f(y_{\alpha\gamma})~,
\end{equation}
where $y_{\alpha\gamma}\equiv\frac{\hbar^2 \beta}{2\mu_{\alpha\gamma}a_{\alpha\gamma}^2}$, 
with $a_{\alpha\gamma}$ the scattering length between the species 
$\alpha$ and $\gamma$. The function $f$ is defined in Eq. (\ref{f(y)}) 
and at the unitary limit $f=1$; $\bar{\omega}_\alpha$ is defined 
in Eq. (\ref{omega_bar}) and we also used the notation
\begin{equation}
\bar{\omega}_{\alpha\gamma}\equiv\prod_{i=x,y,z}\left(\frac{m_\alpha\omega_{\alpha i}^2+
m_\gamma\omega_{\gamma i}^2}{m_{\alpha\gamma}}\right)^{1/6}~.
\end{equation}
Notice that repeating the derivation including Fermi statistics 
one would still get the equations \Eqs{eq:xpxcollave}, 
the only modification being in the temperature dependence 
of the parameter 
$\tau_{\alpha\gamma}$.

Equations \ref{eq:twocomp}, together with \Eqs{eq:xpxcollave}, form a closed set 
of differential equations describing the small oscillations 
of the centers of mass 
of a multi-component fermionic mixture at $T \gtrsim 0.5\max\{T_{F,\alpha}\}$.

\subsection{Two components with equal trapping frequencies}
\label{IIA}

As a simple application of 
\Eqs{eq:twocomp} we retrieve in this Section the well-known case 
of a two-component 
mixture having equal trapping frequencies along the dipole motion direction: 
$\omega_{1 x}=\omega_{2 x}\equiv\omega_x$. 
One easily 
checks that the coordinate and velocity of the center of mass of the mixture 
decouple from the relative position and velocity 
and oscillate with frequency $\omega_x$, according to Kohn theorem. 
By defining 
the (average) relative position and velocity respectively as 
$x_{\textrm{rel}}\equiv\left\langle x \right\rangle_1- \left\langle x \right\rangle_2$ 
and $v_{\textrm{rel}}\equiv\left\langle p_x \right\rangle_1/m_1-
\left\langle p_x \right\rangle_2/m_2$ one gets
\begin{equation*}
\left\{
\begin{array}{l} 
\displaystyle{\frac{dx_{\textrm{rel}}}{dt} -v_{\textrm{rel}}=0}
\\
\displaystyle{\frac{dv_{\textrm{rel}}}{dt}+\omega_x^2 x_{\textrm{rel}} =
-\frac{1}{\tau^\prime} v_{\textrm{rel}}}
\end{array}
\right.
\end{equation*}
where the inverse relaxation time is $\frac{1}{\tau^\prime}=\frac{M_1+M_2}{m_{12}}\frac{1}{\tau_{12}}$ and
$M_\alpha\equiv N_\alpha m_\alpha$ is the total mass of the 
component $\alpha$.
It follows that if 
\begin{equation}\label{crit_cond_2_equal}
\frac{1}{\tau^\prime}>2\omega_x~
\end{equation}
then the relative motion is overdamped: 
as long as the interactions are larger than a critical 
value the transport of momentum is so effective 
that the two components cannot move independently.

\section{Two components with different trapping frequencies}

In this Section we consider a mixture of two components 
in potentials having different trapping frequencies 
($\omega_{1 x},\omega_{1 y},\omega_{1 z})$ and 
($\omega_{2 x},\omega_{2 y},\omega_{2 z})$. 
This case deserves a separate discussion since 
the properties of the solutions of the small oscillation equations 
are similar to those of a three-component mixture
with equal trapping frequencies: 
when the trapping frequencies are equal
along the direction of the dipole motion ($\omega_{1 x}=\omega_{2 x}$), 
two of the eigenvalues of the $4 \times 4$ small oscillation matrix 
are purely imaginary 
due to the Kohn theorem \cite{KOHN61,PITAEVSKII03,PETHICK08}, while for different 
trapping frequencies 
($\omega_{1 x} \neq \omega_{2 x}$)
one can have either
$0$ or $2$ or $4$ real solutions, as it happens for a three-component mixture. 
In this Section we also apply the Sturm theorem 
to count the number of real eigenvalues of the small oscillation matrix 
and we discuss how to characterize the solutions according to this number
starting from the experimental quantities. We finally discuss the 
properties of the Fourier and Laplace transforms 
of the center of mass positions giving informations respectively 
on the imaginary and real part of the eigenvalues.

\begin{figure}[t]
\begin{center}
\includegraphics[width=7.cm,angle=270,clip]{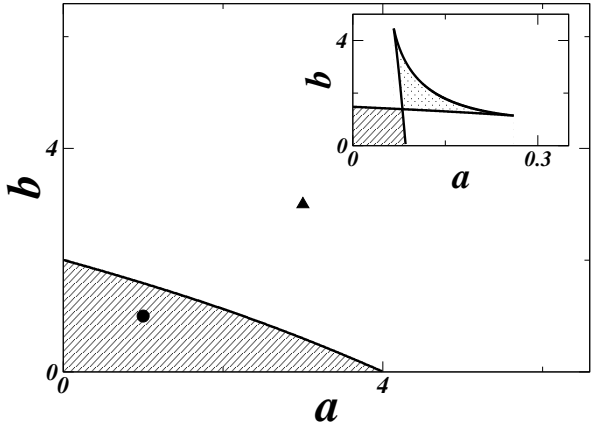} 
\caption{Plot of the regions having $0$ (dashed) and $2$ (blank) real roots 
in the plane $a,b$ ($a=\frac{M_1}{m_{12}}\frac{1}{\tau \omega_{1 x}}$, 
$b=\frac{M_2}{m_{12}}\frac{1}{\tau \omega_{1 x}}$) 
for the case $c=\frac{\omega_{2x}^2}{\omega_{1x}^2}=4$. 
Inset: plot of regions having $0$ (dashed), $2$ (blank) and $4$ (dotted) 
real roots for the case $c= 0.001$.}
\label{due_sturm}
\end{center}
\end{figure} 

\begin{figure}[t]
\begin{center}
\includegraphics[width=7.cm,angle=270,clip]{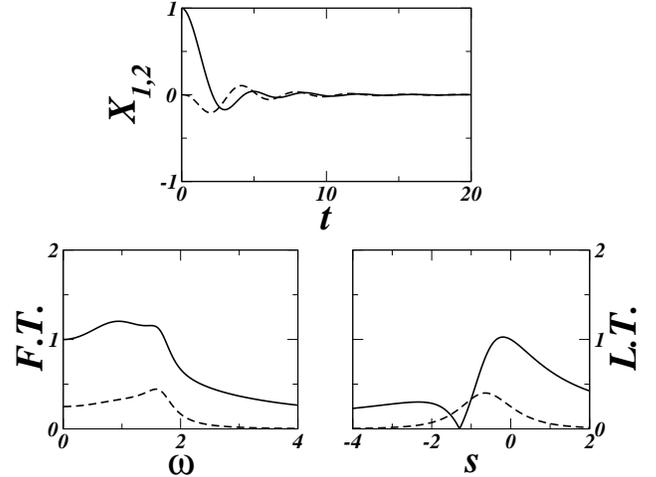} 
\caption{Top: Plot of the center of mass $X_1$ (solid line) and $X_2$ 
(dashed line) as a function of time for $a=1$, $b=1$ (circle 
in Fig.\ref{due_sturm}). Bottom left: modulus of the 
Fourier transform of $X_{1,2}$, 
given by (\ref{Fourier}). 
Bottom right: modulus of the Laplace transform of $X_{1,2}$ defined in (\ref{Laplace_poles}). 
Initial conditions: $X_1(0)=1$, $X_2(0)=0$, 
$V_1(0)=V_2(0)=0$; $c=4$ as in Fig.\ref{due_sturm}.}
\label{due_sturm_0}
\end{center}
\end{figure} 

We define the dimensionless quantities  
$X_\alpha(t)\equiv \left\langle x\right\rangle_\alpha(t)/ L $ and 
$V_\alpha(t)\equiv \left\langle p_x \right\rangle_\alpha(t)/(m_\alpha\omega_{1 x} L)$
($\alpha=1,2$ and 
$L$ is a length scale of the system which can be set equal to an 
harmonic oscillator length); time will be scaled in units of $1/\omega_{1 x}$. 
The equations of motion for the dimensionless 
centers of mass 
positions $X_\alpha$ and velocities $V_\alpha$
are then
\begin{equation}\label{eq:4_eq}
\left\{
\begin{array}{l} 
\dot X_1 =  V_1
\\
\dot X_2 =  V_2
\\
\dot V_1 =  -X_1 - b \left( V_1 - V_2 \right)
\\
\dot V_2 =  -c X_2 + a \left( V_1 
- V_2 \right)~,
\end{array}
\right.
\end{equation}
where 
$a\equiv\frac{M_1}{m_{12}}\frac{1}{\tau \omega_{1 x}}$, 
$b\equiv\frac{M_2}{m_{12}}\frac{1}{\tau \omega_{1 x}}$, 
$c\equiv\frac{\omega_{2 x}^2}{\omega_{1 x}^2}$: 
the parameter $\tau$ is defined in 
Eq. (\ref{eq:tau}) and $M_\alpha\equiv N_\alpha m_\alpha$ is the total mass of the 
component $\alpha$.

It is convenient to introduce a matrix notation for the equations of motion: 
defining the vector $\mathbf{Y} \equiv (X_1,X_2,V_1,V_2)^T $ we 
can write Eqs. (\ref{eq:4_eq}) as 
\begin{equation} \label{matrix_not}
\mathbf{\dot Y} = G_{2}\mathbf{Y}~,
\end{equation} 
where the $4\times 4$ matrix $G_{2}$ is given by
\begin{equation} \label{g2}
G_{2} = 
\begin{pmatrix}
0  & 0  & 1                                         & 0 \\
0  & 0  & 0                                         & 1 \\
-1 & 0  & -b & \phantom{-}b \\
0  & -c & \phantom{-}a & -a \\
\end{pmatrix}
\end{equation}
Of course, once that the eigenvalues $\lambda_1,\cdots,\lambda_4$ 
of the matrix (\ref{g2}) have been determined, it is possible 
to obtain the 
time evolution of the centers of mass. If ${\cal B}$ is the matrix 
diagonalizing $G_2$ such that ${\cal B}^{-1} G_2 {\cal B}=diag\left( 
\lambda_1,\cdots,\lambda_4\right)$, one has
\begin{equation} \label{evol_temp}
\mathbf{Y}_j(t)=\sum_{k=1}^{4}c_{jk}e^{\lambda_k t}
\end{equation}
where the coefficients $c_{jk}$ depend 
upon the initial conditions through the relation 
$c_{jk}={\cal B}_{jk}\sum_{\ell=1}^4 {\cal B}_{k\ell}^{-1} \mathbf{Y}_\ell(0)$.

The properties of the solutions of Eqs. (\ref{matrix_not}) are 
completely determined by the properties 
of the eigenvalues $\lambda_k$ of the matrix (\ref{g2}), 
which are given by the roots of the equation 
\begin{equation} \label{2component}
P(\lambda) = \lambda^4 + (a+b)\lambda^3 + (1+c)\lambda^2 + (a+b c)\lambda+c=0~,
\end{equation}
where $P(\lambda)$ is the characteristic polynomial of the matrix $G_2$. 
Since $P(\lambda)$ is a polynomial with real coefficients, if it has 
a complex root it has also as a root its complex conjugate: this means 
that Eq. (\ref{2component}) has either $0$ or $2$ or $4$ real solutions. 
It is easy to verify that the solutions of the equation $P(\lambda)=0$ 
satisfy the property 
\begin{equation}\label{condition}
\lambda_k^{(R)} \le 0~,
\end{equation}
which just states the stability of solutions (see Appendix B): 
the real parts of the eigenvalues $\lambda_k$ 
correspond to the damping of the normal 
modes and the imaginary parts to their frequencies. 
In Eq. (\ref{condition}), as well as in the following, we denote 
the real part and the imaginary part of $\lambda_k$ respectively 
by $\lambda_k^{(R)}$ and $\lambda_k^{(I)}$. 

One has that $\lambda_k^{(R)}=0$ if and only if $c=1$ 
(i.e., $\omega_{1x} =\omega_{2x}$): indeed if $c=1$, 
$P(\lambda)=(\lambda^2+1)[\lambda^2+(a+b)\lambda+1)]$ and 
two solutions are just $\pm i$. 
These two roots correspond to the oscillations of the center of mass 
of the mixture with frequency equal to the trap one and 
without damping, as discussed in Section \ref{IIA}. Then 
for equal trapping frequencies Eq. (\ref{2component}) can have only 
$0$ or $2$ real solutions, and the critical value for passing from 
one region to the other is just given by Eq. (\ref{crit_cond_2_equal}).

To compute the number of real solutions 
we could look at the general solution of the fourth degree equation 
(\ref{2component}); 
however, in order to generalize this approach to three (or more) components 
it is more convenient and straightforward 
to use the Sturm algorithm \cite{BASU03}, 
which we briefly recall in Appendix B. 
In Fig.\ref{due_sturm} we plot the regions with $0$, $2$ and $4$ real 
roots in the plane $a,b$ for two values of $c$; 
the inset shows that in some narrow region of the 
parameter space ($c=0.001$ in the inset) regions with $4$ 
real roots may occur.

If we adopt the complementary point of view of counting the number of 
independent oscillations (the number of pairs of complex solutions) 
we see that:
only for equal trapping frequencies ($c=1$) we have 
an undamped oscillatory mode (Kohn mode);
on the contrary, for unequal frequencies ($c\neq 1$), 
all the modes are either damped or overdamped. The number of the damped 
oscillatory modes is $2$, $1$ or $0$, corresponding then 
to $0$, $2$ or even $4$ real roots of Eq. (\ref{2component}).

{\it Fourier and Laplace transforms.}
We now discuss how to determine 
the regions with a different number of real roots 
from experimentally measured quantities. 
When the centers of mass of the two components have been determined or 
measured, informations 
on the real and imaginary parts of the eigenvalues $\lambda_k$ can be 
obtained respectively from their Fourier 
and Laplace transforms:
\begin{equation}
F_\alpha(\omega) \equiv \int_0^\infty e^{-i\omega t} X_\alpha(t) dt~,
\label{Fourier}
\end{equation}
\begin{equation}
L_\alpha(s) \equiv \int_0^\infty e^{-st} X_\alpha(t) dt~.
\label{Laplace}
\end{equation}

Each pair of complex roots, say 
$\lambda_1=\lambda_1^{(R)}+i\lambda_1^{(I)}$ and 
$\lambda_2=\lambda_1^{(R)}-i\lambda_1^{(I)}$, 
having an imaginary part $\pm i \lambda_1^{(I)}$, 
should correspond to a maximum of the 
modulus of the Fourier transform 
at $\omega=\lambda_k^{(I)}$: then, one could expect that if the 
Fourier transform has (for positive values of $\omega$) 
two distinct peaks there should be 
$4$ complex eigenvalues (i.e., $0$ real solutions). 
However, Fourier transforms 
of the centers of mass can be analytically computed for arbitrary initial 
conditions, showing that two peaks (at two positive 
values of $\omega$) do not always occur when Eq. (\ref{2component}) 
has $4$ complex roots: e.g., 
if the positions at which the two peaks should be are too close, the peaks
merge into one. Therefore one cannot easily 
infer the number of complex solutions 
from the peaks of the Fourier transforms. 

On the other side, to each real solution corresponds a pole 
in the Laplace transform when $s$ is extended to negative values: 
indeed, setting  $X_\alpha(t)\equiv \sum_{k=1}^4 C_k e^{\lambda_k t}$  
with the coefficients $C_k$ depending upon the initial conditions, 
one has for $s>0$ the analytical expression
\begin{equation}
L_\alpha^{(an.)}(s)=\sum_{k=1}^4 \frac{C_k}{s-\lambda_k}~.
\label{Laplace_poles}
\end{equation}
Eq. (\ref{Laplace_poles}) can be defined also for negative values $s$ and 
it has a pole when $s$ is equal to a real (and negative) eigenvalue.

One can experimentally measure 
$X_\alpha(t)$: of course, to determine the eigenvalues $\lambda_k$ one can
directly fit the data with the expression 
$X_\alpha(t)=\sum_{k=1}^4 C_k e^{\lambda_k t}$. In particular, since 
the number of real eigenvalues is $0$, $2$ or $4$, one can 
use as fitting functions the sum of $4$ exponentials (corresponding to 
$4$ real eigenvalues), or the sum of $2$ exponentials and an exponential 
modulated by a sinusoidal (corresponding to 
$2$ real and $2$ complex conjugate eigenvalues), 
or the sum of $2$ exponentials modulated by 
sinusoidals (corresponding to $4$ complex eigenvalues, i.e. $2$ pairs 
of complex conjugate eigenvalues): the best fit among them 
would determine the number of real eigenvalues. 
Another (possibly complementary) method to determine the eigenvalues 
$\lambda_k$ is to extend 
the definition (\ref{Laplace}) for $L_\alpha(s)$ to negative values of $s$. 
One has that the integral in (\ref{Laplace}) is defined for 
$s>s_{min}=-\min{\{\mid \lambda_k^{(R)} \mid\}}$. Therefore the Laplace 
transform of the center of mass position has a divergence at 
$s=s_{min}$: if 
$\lim_{s\to s_{min}}(s-s_{min})L_\alpha(s)=0$, then $s_{min}$ is the real part 
of a pair of complex conjugate roots (say, $\lambda_1$ and $\lambda_2$) 
and one has to fit from data the corresponding imaginary part $\lambda_1^{(I)}$ 
(with $\lambda_2^{(I)}=-\lambda_1^{(I)}$) and 
coefficients $C_{1,2}$; if not, then 
$s_{min}$ is a real root (say, $\lambda_1$) and 
$\lim_{s\to s_{min}}(s-s_{min})L_\alpha(s)=C_1$. In the latter case, 
one can define a quantity $\tilde{X}_{\alpha}(t) \equiv X_\alpha(t)-C_1 
e^{\lambda_1 t}$: again, from the divergence 
of the Laplace transform of $\tilde{X}_{\alpha}(t)$ 
for negative values of $s$ 
one can determine the next eigenvalue [similarly, 
for a complex root one has to define 
$\tilde{X}_{\alpha}(t) \equiv X_\alpha(t)-C_1 
e^{\lambda_1 t}-C_2 e^{\lambda_2 t}$].   

\begin{figure}[t]
\begin{center}
\includegraphics[width=7.cm,angle=270,clip]{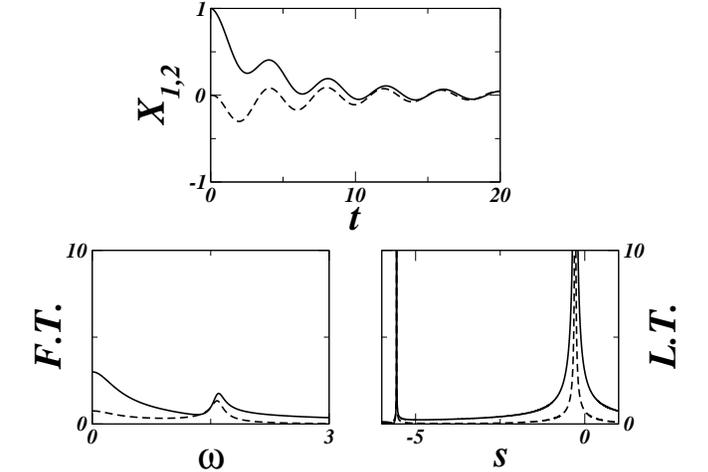} 
\caption{Top: Plot of the center of mass $X_1$ (solid line) and $X_2$ 
(dashed line) as a function of time for $a=3$, $b=3$ (triangle 
in Fig.\ref{due_sturm}). Bottom left: modulus of the Fourier transform of $X_{1,2}$. 
Bottom right: modulus of the Laplace transform of $X_{1,2}$ defined in (\ref{Laplace_poles}). 
Initial conditions: $X_1(0)=1$, $X_2(0)=0$, 
$V_1(0)=V_2(0)=0$; $c=4$ as in Fig.\ref{due_sturm}.}
\label{due_sturm_2}
\end{center}
\end{figure}

The resulting behaviour of the centers of mass positions and 
of the modulus of the Fourier transform (\ref{Fourier}) 
and the Laplace transform (\ref{Laplace_poles}) is plotted 
in Fig.\ref{due_sturm_0} and Fig.\ref{due_sturm_2} 
for two sets of parameters, one belonging to 
the region of zero and one in the region of two real eigenvalues: 
the Laplace transforms show respectively zero and two sharp peaks.

\section{Three components}

In this Section we discuss the properties of the modes 
of three-component mixtures: we consider three-component 
fermionic mixtures of $^{6}Li$, as in the setup reported in 
\cite{OTTENSTEIN08}, and mixtures with two $^{6}Li$ and one $^{40}K$ species 
\cite{WILLE08,SPIEGELHADER09}. We focus for sake of simplicity on 
isotropic potentials ${\cal V}_\alpha$ with the same 
trapping frequency $\omega$. 
Rescaling the variables as in Section III, measuring time in units  
of $1/\omega$ and defining the vector 
$\mathbf{Y} =(X_1,X_2,X_3,V_1,V_2,V_3)^T$ we have $\mathbf{ \dot Y} = 
G_3\mathbf{Y}$, where the $6\times 6$ matrix $G_3$ is:
\begin{eqnarray} \label{g3}
G_3 =
\begin{pmatrix}
    &  &          &     &         &  \\
    & \mathbf{0}  &         &    & \openone & \\
    &   &         &     &  & \\
   &   &        & -\Gamma_{12}-\Gamma_{13} & \Gamma_{12} & \Gamma_{13} \\
   & -\openone  &         & \Gamma_{21} & -\Gamma_{21}-\Gamma_{23} & \Gamma_{23} \\
   &    &         & \Gamma_{31} & \Gamma_{32} & -\Gamma_{31}-\Gamma_{32} \\
\end{pmatrix}&&\nonumber\\
&&
\end{eqnarray}
where $\mathbf{0}$ and $\openone$ are the 
$3 \times 3$ zero and identity matrices and 
$\Gamma_{\alpha \gamma} \equiv \frac{M_\gamma}{m_{\alpha \gamma}} 
\frac{1}{\omega\tau_{\alpha \gamma}}(1-\delta_{\alpha \gamma})$ 
($\alpha,\gamma=1,2,3$) with $\delta_{\alpha\gamma}$ the Kronecker delta. 
The six non-zero parameters $\Gamma_{\alpha\gamma}$ are not independent: 
they satisfy the relation $\Gamma_{12}\Gamma_{23}\Gamma_{31}=
\Gamma_{13}\Gamma_{32}\Gamma_{21}$.
The characteristic polynomial is:
\begin{widetext}
\begin{equation} \label{3component}
P(\lambda) = (\lambda^2 + 1)\Bigg[\lambda^4 + \lambda^3 \sum_{\alpha,\gamma}
\Gamma_{\alpha \gamma} 
+ \lambda^2 \Big(2+ \sum_{\alpha,\gamma,\epsilon\neq\alpha}\Gamma_{\alpha \gamma}
\Gamma_{\gamma \epsilon}+ 
\frac{1}{2}\sum_{\alpha,\gamma,\epsilon\neq\alpha}\Gamma_{\alpha \gamma}\Gamma_{\epsilon\gamma}\Big) + 
\lambda \sum_{\alpha,\gamma}\Gamma_{\alpha \gamma} + 1\Bigg]~.
\end{equation}
\end{widetext}
The discussion proceeds now as in Section III: one has 
\begin{itemize}
\item $\lambda_k^{(R)} \le 0$ (Stability of solutions)
\item $\lambda_k^{(R)} =0$ $\Leftrightarrow$ $\lambda_k^{(I)}=\pm 1$~,
\end{itemize}
the number of real solutions can be either $0$, $2$ or $4$ and the 
number of real roots can be inferred from the 
Laplace transforms of experimental measured quantities. 
At the transition line between one region with $0$ or $2$ real eigenvalues 
and a region having more real roots, $\delta$ functions 
appear and the corresponding strengths 
of these $\delta$ functions 
become non-vanishing. 
The eigenvalues $\lambda=\pm i$ correspond to the center of mass 
oscillation with the frequency of the trap (Kohn theorem). Again, with 
different trapping frequencies, regions with $6$ real eigenvalues
(that is, with no oscillatory mode at all) may occur.

\subsection{Three-component mixture of the same species}

We discuss in this Section the behaviour of the eigenvalues 
of a mixture in which the components
belong to the same species, as in the three-component 
$^{6}Li$ mixture described in \cite{OTTENSTEIN08}. 

For an unpolarized mixture in which each species 
contains an equal number of particles $N_{Li}$ 
($3N_{Li}$ is the total number of particles in the mixture), 
there are three independent parameters $\Gamma_{\alpha\gamma}$ 
(since $\Gamma_{\alpha\gamma}=\Gamma_{\gamma\alpha}$)
and 
the matrix $G_3$ reads
\begin{equation} \label{g3equal}
G_3 = 
\begin{pmatrix}
    &  &          &     &         &  \\
    & \mathbf{0}  &         &    & \openone & \\
    &   &         &     &  & \\
   &   &        & -(\xi_{1}+\xi_{2}) & \xi_{1} & \xi_{2} \\
   & -\openone  &         & \xi_{1} & -(\xi_{1}+\xi_{3}) & \xi_{3} \\
   &    &         & \xi_{2} & \xi_{3} & -(\xi_{2}+\xi_{3}) \\
\end{pmatrix}
\end{equation}
where $\xi_1 \equiv \Gamma_{12}$, $\xi_2 \equiv \Gamma_{13}$ and
$\xi_3 \equiv \Gamma_{23}$. 
The characteristic equation and its solutions are studied in Appendix B.

In Fig.\ref{tre_sturm} we plot the results of our analysis with 
the Sturm theorem in the plane $\xi_1$-$\xi_2$ 
for two values of $\xi_3$. 
If $\xi_3<1$,  
then one has three different 
regions corresponding to $0$, $2$ and $4$ real solutions. 
When $\xi_3>1$  
the region with $0$ real 
solutions disappears 
and the motion is ruled either by one or two oscillating frequencies, but not
three.
In Fig.\ref{tre_sturm_0} the center of mass positions for a simple initial 
condition and the modulus of their Fourier and Laplace (\ref{Laplace_poles}) 
transforms are plotted 
for three sets of parameters, belonging to regions with respectively 
$0$, $2$ and $4$ real eigenvalues showing that 
the Laplace transforms display respectively zero, two and four sharp peaks.
\begin{figure}[t]
\begin{center}
\includegraphics[width=7.cm,angle=270,clip]{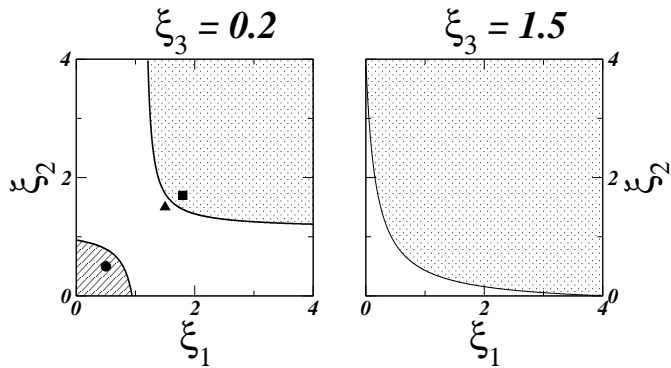} 
\caption{Plot in the $\xi_1$-$\xi_2$ plane of regions having $0$ (dashed), 
$2$ (blank) and $4$ (dotted) real roots for a three-component mixture 
having $\xi_3=0.2$ (left) and $\xi_3=1.5$ (right) - 
the value of $\xi_3$ at which the region with $0$ real roots 
disappears is $\xi_3=1$. The points denoted in the left plot 
are at $\xi_1=0.5$, $\xi_2=0.5$ (circle), $\xi_1=1.5$, $\xi_2=1.5$ 
(triangle), $\xi_1=1.8$, $\xi_2=1.7$ (square).}
\label{tre_sturm}
\end{center}
\end{figure} 
\begin{figure}[t]
\begin{center}
\includegraphics[width=7.cm,angle=270,clip]{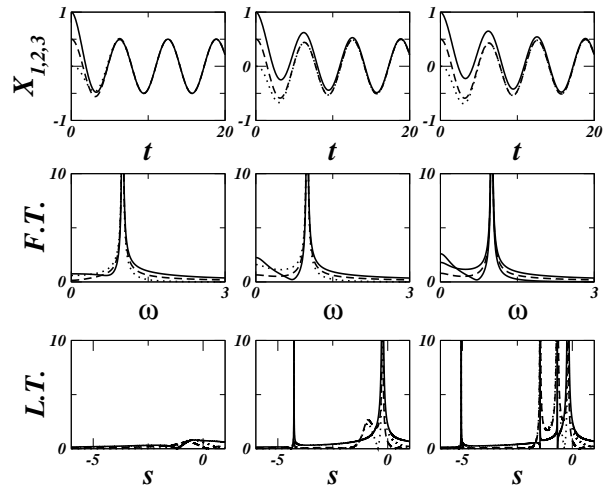} 
\caption{Plot of the center of mass $X_{1,2,3}(t)$ (top row), 
modulus of the Fourier transform $F_{1,2,3}(\omega)$ (central row) and 
modulus of the Laplace transform $L_{1,2,3}(s)$ defined in 
(\ref{Laplace_poles}) (bottom row). 
The left, central and right columns correspond respectively 
to the point $\xi_1=0.5$, $\xi_2=0.5$ (circle in Fig.\ref{tre_sturm}) - 
$0$ real roots), 
$\xi_1=1.5$, $\xi_2=1.5$ (triangle - $2$ real roots), 
$\xi_1=1.8$, $\xi_2=1.7$ (square - $4$ real roots). 
Solid, dashed and dotted lines are for the corresponding quantities 
of the species $1$, $2$ and $3$. Initial conditions: $X_1(0)=1$, $X_2(0)=0.5$, 
$X_3(0)=0$, $V_1(0)=V_2(0)=V_3(0)=0$.}
\label{tre_sturm_0}
\end{center}
\end{figure} 

\begin{figure}[t]
\begin{center}
\includegraphics[width=7.cm,angle=270,clip]{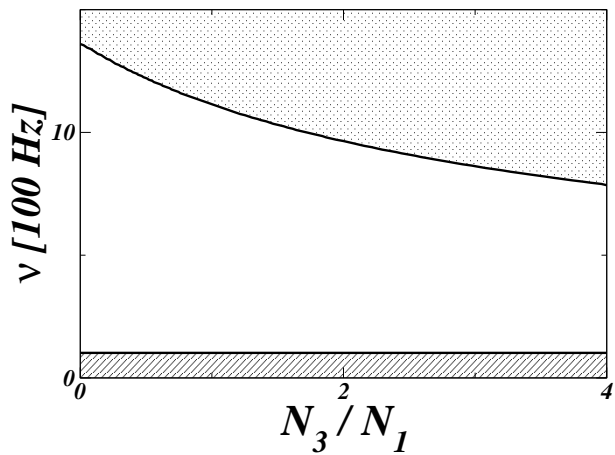} 
\caption{Plot of regions having $0$ (dashed), $2$ (blank) and $4$ (dotted) real roots  
for a mixture of three $Li$ species when the number 
$N_3$ and the trapping frequency 
$\omega=2\pi \nu$ are varied. The components $1$ and $2$ are taken at unitary limit 
with a fixed number of atoms $N_1=N_2=5 \cdot 10^4$. 
Temperature is fixed ($T=500nK$) and 
the two remaining scattering lengths are chosen 
to be $a_{13}=a_{23}=400a_0$ (similar plots are obtained 
when $a_{13}$ and $a_{23}$ are different, but small).}
\label{tre_sturm_2}
\end{center}
\end{figure}

When the mixture is polarized, one has to use \Eq{g3}:
varying the number 
of the atoms in the components, it is possible to explore the different regions 
through the polarization. An example is shown in Fig.\ref{tre_sturm_2}, 
where the number of atoms of the component $3$ is varied and the components 
$1$ and $2$ are at the unitary limit. As discussed in \cite{OTTENSTEIN08}, 
the mixture is collisionally stable for small values of the other scattering 
lengths. Fig.\ref{tre_sturm_2} shows that by varying the trapping frequency one 
can explore the three regions with $0$, $2$ and $4$ real eigenvalues.
Notice that the stronger the confinement due to the trap, the fewer independent 
oscillatory modes are possible.

\subsection{Different species}

In this Section we present the results for a three-component mixture of 
two species of $^{6}Li$ (denoted by $Li_1$ and $Li_2$) and one species 
of $^{40}K$ recently realized experimentally: a small sample 
of $^{40}K$ is immersed in a two-component $^6Li$ mixture and 
found to be collisionally stable with 
very low loss rates for large negative scattering lengths between the two 
$^6Li$ (for $\mid 
a_{Li-Li} \mid \gtrsim 1400 a_0$, where $a_0$ is the Bohr radius) 
\cite{SPIEGELHADER09}. 
The matrix $G_3$ of \Eq{g3} simplifies to
\begin{equation} \label{g3different}
G_3 = 
\begin{pmatrix}
    &  &          &     &         &  \\
    & \mathbf{0}  &         &    & \openone & \\
    &   &         &     &  & \\
   &   &        & -(\xi_{1}+\xi_{2} r \alpha) & \xi_{1} & \xi_{2} r \alpha \\
   & -\openone  &         & \xi_{1} & -(\xi_{1}+\xi_{2} r \alpha) & \xi_{2} r \alpha \\
   &    &         & \xi_{2} & \xi_{2} & -2\xi_{2} \\
\end{pmatrix}
\end{equation}
where we defined 
$\xi_1 \equiv \Gamma_{12}$, $\xi_2 \equiv \Gamma_{31}$, 
$r\equiv m_{K}/m_{Li}$ and $\alpha\equiv N_K / N_{Li}$. 
To simplify formulas, we assumed the same number of atoms in the two Li species and 
equal isotropic
trapping frequencies, but similar 
(even if more involved) expressions can be found for the realistic case 
of anisotropic cigar-shaped potentials and with the trapping frequencies 
of $Li_1$-$Li_2$ different from those of the $K$ component: 
they will be reported elsewhere.

\begin{figure}[t]
\begin{center}
\includegraphics[width=7.cm,angle=270,clip]{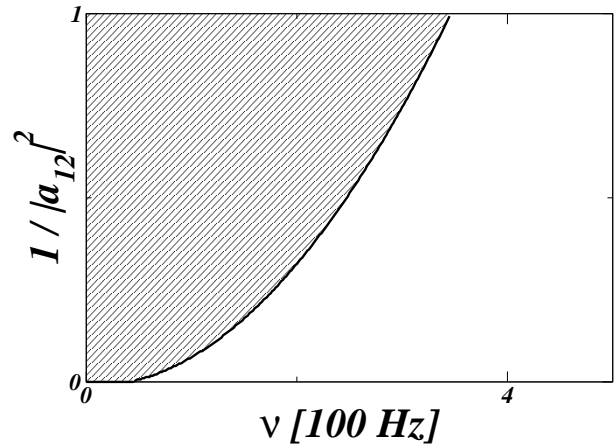} 
\caption{Plot of regions having $0$ (dashed) and $2$ 
(blank) real roots for a mixture of $Li_1$-$Li_2$-$K$ with 
equal trapping frequencies $\omega=2 \pi \nu$ and equal number of 
$Li$ atoms $N_{Li_1}=N_{Li_2} 
\equiv N_{Li}= 5 \cdot 10^4$. The number 
of $K$ atoms has been chosen $N_K=0.25 N_{Li}$. 
Parameters: $T=200nK$, $a_{Li_1-K}=a_{Li_2-K} = 63 a_0$; the $Li_1-Li_2$ 
scattering length 
$a_{12}$ is measured in units of $100a_0$. 
For $\nu<55Hz$ there are $0$ real roots whatever $a_{12}$ is.}
\label{Li_K}
\end{center}
\end{figure}

In Fig.\ref{Li_K} we plot the number of real solutions 
taking a mixture with the same number of particles 
for the two Lithium components and a smaller number of particles of the 
$K$ component. We vary the scattering length between $Li_1$ and $Li_2$ 
and the
trapping frequency $\omega$. 

The number of oscillatory modes is reduced, again, by increasing the interaction
strength or making the trap more confining.

\section{Conclusions}

In this paper we used the Boltzmann equation to study dipole oscillations 
in a general fermionic mixture: to discuss and classify the different solutions 
in the parameter space through it, we used the number of real 
eigenvalues of the small oscillations matrix, 
discussing how this number is related
to the Laplace transform of the experimental quantities. 
The small oscillation matrix for a $N$-component mixture
has $2N$ eigenvalues
and the complex ones
appear in pairs of complex conjugates: 
the number of independent collective mode 
frequencies
is therefore given by $(2N-\textrm{number of real roots})/2$. Counting the real roots 
(overdamped modes) and the complex ones (oscillatory ones)
are complementary descriptions. The reason why we chose the first
one is to take advantage of Sturm algorithm and Laplace transform properties.

After deriving the equations of motion for the centers of mass 
of the components and discussing as a simple example the 
case of two components 
in harmonic potentials 
having different trapping frequencies, we studied dipole oscillations in 
three-component mixtures.

A $N$-component mixture where each component experiences
a different trapping has a maximum of $N$ independent oscillation frequencies.
We found, as expected, that when interactions are favoured 
(by increasing the strength of interactions
between components, or shrinking the traps) the components tend to move together and
the number of independent 
frequencies reduces.
If all the trap frequencies are equal along one direction
there is one undamped oscillation with frequency equal to the trap one
(Kohn theorem). Beside that,
there can be up to $N-1$ damped oscillatory modes.
On the contrary,
if the trap frequencies are different, there is no purely oscillatory
mode. The number of damped oscillatory modes can be at most $N$ and can be reduced
down to $0$: it is possible that there is no 
collective dipole oscillatory mode at all in the mixture.

In explicit computations we used the Boltzmann equation with classical
collision terms and without intra-species interactions: this 
gives, for strong inter-species interactions, results very close to the findings obtained including
Fermi statistics and in-medium effects thanks to a compensation of effects 
\cite{RIEDL09,CHIACCHIERA09}. 
Using the classical Boltzmann equation 
without intra-species interactions is expected to be a good approximation 
(also far from the unitary limit) at temperatures above
$\sim 0.5 \max\{T_{F,\alpha}\}$, where $T_{F,\alpha}$ is the Fermi temperature of 
the component $\alpha$: for two components, e.g., 
it well reproduces the experimental results for the scissor mode for these 
temperatures \cite{RIEDL09,CHIACCHIERA09}. We also observe 
that the explicit inclusion of the Fermi statistics modifies the dependence 
of the relaxation times upon the system parameters, but not 
the dynamical equations for the centers of mass and their momenta and then  
the subsequent analysis based on 
the Sturm theorem would proceed in the same way.

The focus of this paper has been on the study of dipole oscillations, 
however we point out that a similar study  
can be done for different (and more complicated) collective modes: one has 
to write the small oscillations equations, and determine the number 
of real roots of the small oscillations matrix.

To conclude, we observe that 
multi-component mixtures are promising to study mechanisms of color superconductivity 
\cite{ALFORD08} due to the possibility of having many components 
with tunable interactions: 
from this point of view, to study the collective oscillations 
(and follow the modifications and the fate of the regions in the 
parameter space discussed in this paper) 
when the temperature is lowered until the superfluid critical 
temperature can provide useful informations on possible intermediate, new 
strongly correlated phases.

\section*{ACKNOWLEDGEMENTS} We thank for fruitful discussions M. Urban, 
S. Jochim, F. Schreck and S. Moroni. 
This work is supported by the grants INSTANS (from ESF) and 2007JHLPEZ (from MIUR).

\appendix

\section{Computation of $\tau$}

\begin{figure}[t]
\begin{center}
\includegraphics[width=7.cm,angle=270,clip]{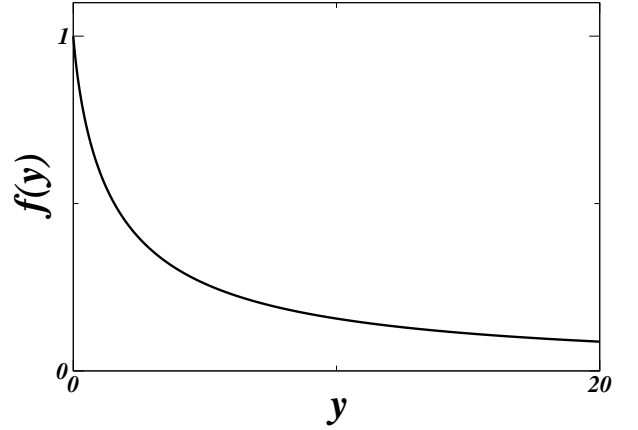} 
\caption{Plot of the function $f(y)$ defined in Eq.(\ref{f(y)}) as a function of the dimensionless parameter 
$y= \hbar^2 \beta / 2\mu a_{12}^2$.}
\label{f_y}
\end{center}
\end{figure}

In this appendix we give some details on
the computation of the parameter 
$\tau$ related to the collisions between atoms  
of species $1$ and atoms of species $2$. 
As usual, given two particles with masses $m_1,m_2$, positions $\vek{r}_1,\vek{r}_2$
and momenta $\vek{p}_1,\vek{p}_2$ the global and relative coordinate and momenta are defined as:
$\vek P \equiv\vek p_1+ \vek p_2$, $\vek R\equiv(m_1\vek r_1+m_2\vek r_2)/m_{12}$, $\vek r_{\textrm{rel}}\equiv
\vek r_1-\vek r_2$, $\vek p_{\textrm{rel}}\equiv(m_2 \vek p_1-m_1\vek p_2)/m_{12}$.
Then $p_1^2/2m_1+p_2^2/2m_2=P^2/2m_{12}+p_{\textrm{rel}}^2/2\mu$ and 
$\vek p_{\textrm{rel}}=\mu \dot{\vek r}_{\textrm{rel}}$, where 
$\mu\equiv\frac{m_1m_2}{m_1+m_2}$ is the reduced mass and 
$m_{12}\equiv m_1+m_2$. 
For simplicity we consider equal isotropic trapping frequencies: 
$\omega_{1,i}=\omega_{2,i} \equiv \omega$ ($i=x,y,z$); 
$a_{12}$ is the scattering length and 
$\frac{d\sigma}{d\Omega}= \frac{1}{p_{\textrm{rel}}^2/\hbar^2 + 1/a_{12}^2}$ 
the differential cross section.
The collisional average $\left\langle I_{12} p_x \right\rangle_{ 1}$ 
is defined as
\begin{equation}\label{app_coll_aver}
\left\langle I_{12} p_x \right\rangle_{ 1} = 
\frac{1}{N_1} \int d\Gamma I_{12} p_x~.
\end{equation}
Plugging the Gaussian ansatz for the distribution function 
(\ref{eq:gaussianansatz}) in Eq. (\ref{app_coll_aver}) 
and moving to the center of mass and relative coordinates 
we obtain
\begin{equation}
\left\langle I_{12} p_x \right\rangle_{ 1} =  \frac{2\beta^4 \omega^3}{3 \pi} 
\frac{N_2 v_{\textrm{rel}}}{\mu} \int_0^\infty dp_{\textrm{rel}} 
\frac{p_{\textrm{rel}}^5}{\frac{p_{\textrm{rel}}^2}{\hbar^2} + \frac{1}{a_{12}^2}} e^\frac{-\beta p_{\textrm{rel}}^2}{2\mu}~
\end{equation}
where $\mu$ is the reduced mass and $v_{\textrm{rel}}$ the relative velocity
between the two components. 
Setting $\frac{\beta}{2\mu}p_{\textrm{rel}}^2 = t$ then we have:
\begin{equation}\label{app_integral}
\left\langle I_{12} p_x \right\rangle_{ 1}  = 
\frac{4\hbar^2 \beta^2 \omega^3}{3 \pi} N_2 v_{\textrm{rel}} 
\mu \int_0^\infty dt \frac{t^2}{t+y}e^{-t}~
\end{equation}
where $y=\frac{\hbar^2 \beta}{2\mu a_{12}^2}$. 
The last integral in the right-hand side of (\ref{app_integral}) can be written 
in terms of the generalized $\Gamma$-function 
$\Gamma(0,x) = \int_x^\infty dt \frac{e^{-t}}{t}$ \cite{ABRAMOWITZ72} as
\begin{equation} \label{f(y)}
f(y) = \int_0^\infty dt \frac{t^2}{t+ y} e^{-t} = 1-y+ y^2e^y \Gamma(0,y)~.
\end{equation}
The function $f(y)$ is plotted in Fig.\ref{f_y}: it is $f(0)=1$ 
(at the unitary limit) and $f$ vanishing for large $y$.  
One then obtains
\begin{equation}
\frac{1}{\tau} = \frac{4\beta^2 \hbar^2 \omega^3}{3\pi} f(y)~.
\end{equation}
The computation for general anisotropic and different 
trapping frequencies can be similarly done and gives the result 
(\ref{eq:tau}).

\section{Sturm chains and Sturm theorem.}

Here we recall how to apply Sturm theorem to count the number 
of real roots of a polynomial equation $P(\lambda)=0$ \cite{BASU03}.
First, one has to 
define the (finite) sequence of polynomials (\textit{Sturm chain}) as follows:
\begin{eqnarray} \label{eq:sturmchain}
P_0(\lambda) &\equiv&  P(\lambda)\nonumber\\
P_1(\lambda) &\equiv& P^\prime(\lambda)\nonumber\\
P_2(\lambda) &\equiv& - \text{rem}(P_0(\lambda),P_1(\lambda))\nonumber\\
P_3(\lambda) &\equiv& - \text{rem}(P_1(\lambda),P_2(\lambda))\nonumber\\
\cdots  &\equiv & \cdots\nonumber\\
0         &=  &- \text{rem}(P_{m-1}(\lambda),P_m(\lambda))~,\nonumber\\
\end{eqnarray}
where we denote with $\text{rem}(P_{m-1}(\lambda),P_m(\lambda))$ 
the rest of the polynomial division of the polynomials $P_{m-1}(\lambda)$ 
and $P_m(\lambda)$. Then one has to evaluate the sign of the polynomials 
$P_i(\lambda)$ at $+\infty$ and $-\infty$ and 
to compute the number of changes of sign 
(which we call $\Delta(+\infty)$ and $\Delta(-\infty)$) 
in the sequences $\{ P_0(+\infty), P_1(+\infty),P_2(+\infty),\dots \}$ 
and $\{ P_0(-\infty), P_1(-\infty),P_2(-\infty),\dots \}$. 
Sturm theorem states that the number of real solutions of the equation 
$P(x)=0$ is given by $|\Delta(+\infty)- \Delta(-\infty)|$. Notice that the 
Sturm theorem is not directly applicable to cases where there are degenerate solutions.
In these cases, one has first to transform the equation into one that has only simple 
roots, but
this can always be done.

We show how to apply Sturm theorem in the specific case of polynomial 
(\ref{2component}) of fourth degree 
\begin{equation}\label{2component-app}
P(\lambda) = \lambda^4 +\lambda^3(a+b) + \lambda^2(1+c) + 
\lambda(a+bc) + c~.
\end{equation}
In order to evaluate the sign of Sturm polynomials at 
$\lambda = \pm \infty$ we determine the sign of the coefficient of the highest power in the polynomial as follows:

\begin{equation*}
\begin{array}{c|ccccc}
 \lambda            &     P_0    &     P_1    &     P_2               &     P_3                &   P_4    \\ \hline
+\infty      &       +      &     +         &     (-1)^\alpha    &     \phantom{-} (-1)^\beta       &   (-1)^\gamma \\ \hline
-\infty      &       +      &     -         &     (-1)^\alpha    &     -(-1)^\beta       &   (-1)^\gamma \\ 
\end{array}
\end{equation*}
where $\alpha,\beta, \gamma = \{ 0,1\}$.
Then number of changes of signs in the two cases is:
\begin{eqnarray*}
\Delta(+\infty) &= & \frac{3}{2} - \frac{1}{2}\left[ (-1)^\alpha + (-1)^\alpha(-1)^\beta + (-1)^\beta(-1)^\gamma       \right]\\
\Delta(-\infty)  &= & \frac{5}{2} + \frac{1}{2}\left[ (-1)^\alpha + (-1)^\alpha(-1)^\beta + (-1)^\beta(-1)^\gamma       \right]~
\end{eqnarray*}
and the function counting the number of real solutions is
\begin{equation}  \label{counting}       \sst{
|\Delta(+\infty)- \Delta(-\infty)| = |1 + (-1)^\alpha + (-1)^\alpha(-1)^\beta + (-1)^\beta(-1)^\gamma}|~.
\end{equation}
The counting function (\ref{counting}) is then

\begin{center}
\begin{tabular}{c|cc|cc}
&\multicolumn{2}{c|}{$\alpha$=0}&
\multicolumn{2}{c}{$\alpha$=1}\\
\hline
  &   $\gamma$=0 &   $\gamma$=1 &  $\gamma$=0 &   $\gamma$=1  \\ 
\hline
$\beta$=0  & 4  &  2 & 0 & 2\\
$\beta$=1  & 0  &  2 & 0 & 2\\
\end{tabular} 
\end{center}

When the real part of the root of Eq. (\ref{2component-app}) is negative, 
then solutions of equations of motion (\ref{eq:4_eq}) 
do not grow exponentially: writing $P(\lambda)=
\lambda^4 + A\lambda^3 +B\lambda^2 + C\lambda + D$ 
this happens if \cite{BIRKHOFF77}
\begin{align*} 
& A>0 \ \ B>0 \ \ C>0 \ \ D > 0\\
& ABC > A^2D + C^2.
\end{align*}
It is easy to verify that these conditions are satisfied. 

The characteristic polynomial of the matrix $G_3$ in Eq.(\ref{g3equal}) is:
\begin{align*}  
P(\lambda) = &(\lambda^2+ 1)\{\lambda^4 + 
2\lambda^3( \xi_1 + \xi_2 + \xi_3 ) \\
&+\lambda^2 [2 + 3(\xi_1 \xi_2 +\xi_2 \xi_3 + \xi_1 \xi_3)] \\
           &+ 2 \lambda (\xi_1 + \xi_2 + \xi_3 ) + 1\}.
\end{align*} 

For completeness we write here the explicit 
solutions of the quartic equation $P(\lambda)/(\lambda^2+ 1) =0$:
\begin{eqnarray*}
  \lambda_{I,II} &=& \frac{1}{2}\left(-S-\sqrt{Q_1}\pm\sqrt{-4+Q_2+2 S \sqrt{Q_1}}\right)\\
  \lambda_{III,IV} &=& \frac{1}{2}\left(-S-\sqrt{Q_1}\pm\sqrt{-4+Q_2-2 S \sqrt{Q_1}}\right)~,
\end{eqnarray*}
where 
\begin{eqnarray*}
 S(\xi_1,\xi_2,\xi_3)&\equiv& \xi_1 + \xi_2 + \xi_3\\
 Q_1(\xi_1,\xi_2,\xi_3) &\equiv& \xi^2_1+\xi^2_2+\xi^2_3-\xi_1\xi_2-\xi_2 \xi_3-\xi_1\xi_3\\
 Q_2(\xi_1,\xi_2,\xi_3)& \equiv& 2(\xi_1+\xi_2+\xi_3)^2-3(\xi_1 \xi_2 +\xi_2 \xi_3 + \xi_1 \xi_3)~.
\end{eqnarray*}

\end{document}